\documentclass[a4paper]{jpconf}

\usepackage{graphicx}
\usepackage{amsmath}
\usepackage{amssymb}

\begin{document}
\title{Four approaches for description of stochastic systems with small and finite inertia}

\author{E V Permyakova$^{1,2}$, L S Klimenko$^{1,2}$, I V Tyulkina$^{1,3}$ and D~S~Goldobin$^{1,3}$}
\address{$^1$Institute of Continuous Media Mechanics UB RAS, 1 Akademika Koroleva street,\\
         Perm 614013, Russia}
\address{$^2$Department of Theoretical Physics, Perm State University, 15 Bukireva street,\\
         Perm 614990, Russia}
\address{$^3$Department of Control Theory, Lobachevsky University of Nizhny Novgorod, 23 Gagarin avenue,
         Nizhny Novgorod 606950, Russia}
\ead{Denis.Goldobin@gmail.com}

\begin{abstract}
We analyse four approaches to elimination of a fast variable, which are applicable to systems like passive Brownian particles: (i)~moment formalism, (ii)~corresponding cumulant formalism, (iii)~Hermite function basis, (iv)~formal `cumulants' for the Hermit function basis. The accuracy and its strong order are assessed. The applicability and performance of two first approaches are also demonstrated for active Brownian particles.
\end{abstract}

\section{Introduction}
Characterization of dynamics of overdamped systems can be often reduced to a single variable, which can be coordinate for mechanical systems in a viscous media (like Brownian particles)~\cite{Haken-1977,Gardiner-1983,Becker-1985} or an oscillation phase for self-sustained periodic oscillators~\cite{Winfree-1967,Kuramoto-1975}, where the transversal deviations from the limit cycle decay quick enough to be neglected. However, in stochastic systems with delta-correlated noise, this reduction becomes nontrivial as the inertia term is non-small for rapid fluctuations in mechanical systems~\cite{Haken-1977,Gardiner-1983,Becker-1985,Wilemski-1976,Gardiner-1984,Goldobin-Klimenko-2020} and the deviations from the limit cycle are non-negligible for oscillatory systems~\cite{Yoshimura-Arai-2008,Teramae-etal-2009,Goldobin-etal-2010}. For the phase equation of oscillatory systems, an effective inertia-like term may appear owing to different reasons leading to a significant increase of the dynamics complexity~\cite{Komarov-Gupta-Pikovsky-2014,Olmi-etal-2014,Bountis-etal-2014,Jaros-Maistrenko-Kapitaniak-2015,Zharkov-Altudov-1978}.
The problem of the transition to the small inertia limit, in other words, the problem of adiabatic elimination of a fast variable (velocity), was thoroughly addressed in the literature for passive Brownian particles~\cite{Haken-1977,Gardiner-1983,Becker-1985,Wilemski-1976,Gardiner-1984,Goldobin-Klimenko-2020} and for some types of active Brownian particles~\cite{Milster-etal-2017}.

Recently, a systematic approach to the construction of low-dimensional model reductions for oscillator populations was suggested on the basis of the circular cumulant representation~\cite{Tyulkina-etal-2018,Goldobin-etal-2018,Goldobin-Dolmatova-2019}; this approach generalizes the Ott--Antonsen ansatz~\cite{Ott-Antonsen-2008,Ott-Antonsen-2009} based on the Watanabe--Strogatz partial integrability~\cite{Watanabe-Strogatz-1993,Watanabe-Strogatz-1994,Pikovsky-Rosenblum-2008,Marvel-Mirollo-Strogatz-2009}.
Application of this new approach to systems with non-negligible inertia necessitates a systematic analysis of possible approaches to the problem of elimination of a fast variable. In this paper we provide such an analysis with the focus on non-conventional techniques.

\section{Results}
We consider the Langevin equation with inertia
\begin{equation}
\mu\ddot\varphi+\dot\varphi=F(\varphi,t)+\sigma\xi(t)\,,\qquad\mu\ll1\,,
\label{eq001}
\end{equation}
where $\mu$ is mass or a measure of `inertia' in the system (for Josephson junctions~\cite{Zharkov-Altudov-1978}, power grid models~\cite{Morren-etal-2006,Short-Infield-Freris-2007}, etc.), $F$ is a deterministic force, $\sigma$ is the noise strength, $\xi(t)$ is a normalized $\delta$-correlated Gaussian noise: $\langle\xi\rangle=0$, $\langle\xi(t)\,\xi(t')\rangle=2\delta(t-t')$.

The Fokker--Planck equation for the probability density $\rho(v,\varphi)$, where $v\equiv\dot\varphi$, reads
\begin{equation}
\partial_t\rho=-v\partial_\varphi\rho+\partial_v\left[\frac{1}{\mu}\big(v-F(\varphi,t)\big)\rho\right]
+\frac{\sigma^2}{\mu^2}\partial_v^2\rho
\label{eq002}
\end{equation}
($\varphi$ may be in a rotating reference frame, where it is useful).

Our aim is to eliminate the velocity and consider effective dynamics solely for $\varphi$. We analyse four approaches to accomplishing this task:
\begin{itemize}
\item
Moment formalism: representation in terms of $w_n(\varphi)=\int_{-\infty}^{+\infty}v^n\rho(v,\varphi)\,\mathrm{d}v$\,;
\\
calculations with equations~(\ref{eq003})--(\ref{eq005}) (or equation~(\ref{eq-abp-03}) for active Brownian particles).
\\
Adiabatic elimination requires elements 0--2; the $\mu^1$-correction: 0--4; the $\mu^m$-approximation: 0--$(2m+2)$.
\item
Cumulant formalism: representation in terms of $K_n(\varphi)$ (or $\varkappa_n=\frac{K_n}{n!}$), defined as follows: $f(s,\varphi)=\sum_{n=0}^{\infty}w_n(\varphi)\frac{s^n}{n!}$, $\ln{f}=\phi(s,\varphi)=\sum_{n=0}^{\infty}K_n(\varphi)\frac{s^n}{n!}$~\cite{Lukacs-1970};
\\
calculations with equations~(\ref{eq025})--(\ref{eq026}).
\\
Adiabatic elimination requires elements 0--2; the $\mu^1$-correction: 0--2 (for the adiabatic elimination fewer number of contributions for these elements are included); the $\mu^m$-approximation: 0--$(m+1)$.
\item
Basis of Hermite functions $h_n(u)$, which are the eigenfunctions of operator $\hat{L}_1=\partial_u(u+\partial_u)$:
\[
\textstyle
\rho(v,\varphi)=\sum_{n=0}^\infty \frac{\sigma}{\sqrt{\mu}}\,h_n\!\!\left(\frac{\sqrt{\mu}}{\sigma}v\right)W_n(\varphi,t)\,;
\]
calculations with equations~(\ref{eqH04})--(\ref{eqH05}).
\\
Adiabatic elimination requires elements 0--1; the $\mu^1$-correction: 0--2; the $\mu^m$-approximation: 0--$(m+1)$.
\item
An analogue of the cumulant representation for the basis of Hermite functions:
representation in terms of $\varkappa_n$ defined via
$f(s,\varphi)=\sum_{n=0}^{\infty}W_n(\varphi)s^n$ (notice, no $n!$) and $\ln{f}=\phi(s,\varphi)=\sum_{n=0}^{\infty}\varkappa_n(\varphi)s^n$;
\\
calculations with equations~(\ref{eqH06})--(\ref{eqH07}).
\\
Adiabatic elimination requires elements 0--1; the $\mu^1$-correction: 0--2; the $\mu^m$-approximation: 0--$(m+1)$.
\end{itemize}

The numerical simulations for $F=0.5-1.8\sin\varphi$, which is relevant for the study~\cite{Komarov-Gupta-Pikovsky-2014}, revealed the following.
The actual accuracy of all approximations for a given order of approximation is practically the same. Meanwhile, the behavior of elements with $n$ significantly differs.
%Notice, usual moments $W_n$ growth with $n$ approximately as $\sqrt{n!}$ ($W_n$ are $w_n$ rescaled according to the scaling law with respect to $\mu$).
The most regular scaling with the growth on $n$ is observed for `cumulants' for the Hermite basis. For the plain cumulants, the $\varkappa_2$ is large, as it should be, but the higher-order elements gradually decay. Noticeably, in spite these elements are not as small as the `cumulants' for the Hermite basis, the truncation at the same $m$-th element leads to the same accuracy for the probability density evolution of $\varphi$.

For the case of active Brownian particles~\cite{Milster-etal-2017,Lighthill-1952,Blake-1971,Ebbens-Howse-2010}, one can immediately employ the moment or cumulant formalisms, while the Hermit function basis needs to be significantly corrected. With the latter approach an individual mathematical preparation for each new problem is required, which can be problematic. Generally, calculations with system~(\ref{eq-abp-03}) for active Brownian particles requires a large number of terms in series and might suffer from numerical instabilities. We overcome these challenges by employing a modification~\cite{Permyakova-Goldobin-2020} of the exponential time differencing method~\cite{Cox-Matthews-2002} which provides high performance and accuracy for stiff systems. %~\cite{Matthews-Cox-2000a,Matthews-Cox-2000b}.

\section{Methods}
%\section{Fokker--Planck equation}
\subsection{Moment formalism for Fokker--Planck equation}
One can introduce the moments for $v$:
\[
w_n(\varphi)=\int_{-\infty}^{+\infty}v^n\rho(v,\varphi)\,\mathrm{d}v\,.
\]
For these moments the Fokker--Planck equation~(\ref{eq002}) yields
\begin{align}
&\partial_tw_0+\partial_\varphi w_1=0\,,
\label{eq003}\\
&w_1+\mu\partial_tw_1=F w_0-\mu\partial_\varphi w_2\,,
\label{eq004}\\
&w_n+\frac{\mu}{n}\partial_t w_n=F w_{n-1}-\frac{\mu}{n}\partial_\varphi w_{n+1}
%\nonumber\\
%&\hspace{2cm} {}
 +(n-1)\frac{\sigma^2}{\mu}w_{n-2}\quad\mbox{ for } n\ge2\,.
\label{eq005}
\end{align}

For constructing a regular perturbation theory with respect to small parameter $\mu$ it is convenient to take the scaling laws for $\langle{v^n}\rangle$ with respect to $\mu$~\cite{Goldobin-Klimenko-2020} into account explicitly by means of rescaling
\[
w_n
=\left\{
\begin{array}{cl}
\displaystyle
\frac{1}{\mu^{n/2}}W_n\,,
& \mbox{ for even }$n$\,;\\
\displaystyle
\frac{1}{\mu^{(n-1)/2}}W_n\,,
& \mbox{ for odd }$n$\,.
\end{array}
\right.
\]
Then equations~(\ref{eq003})--(\ref{eq005}) can be recast in a form free from $1/\mu$-coefficients:
%\begin{align}
%&\partial_t W_0+\partial_\varphi W_1=0\,;
%\label{eq006}\\
%&W_1+\mu\partial_t W_1=F W_0-\partial_\varphi W_2\,;
%\label{eq007}\\
%&W_n+\frac{\mu}{n}\partial_t W_n=\mu F W_{n-1}-\frac{\mu}{n}\partial_\varphi W_{n+1}
%\nonumber\\
%&\hspace{1.5cm} {}+(n-1)\sigma^2W_{n-2}\quad\mbox{ for } n=2m\,;
%\label{eq008}\\
%&W_n+\frac{\mu}{n}\partial_t W_n=F W_{n-1}-\frac{1}{n}\partial_\varphi W_{n+1}
%\nonumber\\
%&\hspace{1.5cm} {}+(n-1)\sigma^2W_{n-2}\quad\mbox{ for } n=2m+1\,.
%\label{eq009}
%\end{align}
%Rearranging the terms, one finds
\begin{align}
&\partial_t W_0+\partial_\varphi W_1=0\,,
\label{eq010}\\
&W_1=F W_0-\partial_\varphi W_2-\mu\partial_t W_1\,,
\label{eq011}\\
&W_n=(n-1)\sigma^2W_{n-2}+\mu\Big[F W_{n-1}-\frac{1}{n}\partial_\varphi W_{n+1}
%\nonumber\\
%&\hspace{1.5cm} {}
-\frac{1}{n}\partial_t W_n\Big]\qquad\mbox{ for } n=2m\,,
\label{eq012}\\
&W_n=(n-1)\sigma^2W_{n-2}+F W_{n-1}-\frac{1}{n}\partial_\varphi W_{n+1}
%\nonumber\\
%&\hspace{1.5cm} {}
-\frac{\mu}{n}\partial_t W_n\qquad\mbox{ for } n=2m+1\,.
\label{eq013}
\end{align}
%This equation system contains only $\mu^0$- and $\mu^1$-terms; thus, it is suitable for dealing with the limit $\mu\to0$.

\subsubsection{Elimination of a fast variable}
System~(\ref{eq010})--(\ref{eq013}) for $\mu=0$ yields, after some algebra~\cite{Goldobin-Klimenko-2020},
%\begin{align}
%&\textstyle
%\partial_t W_0+\partial_\varphi W_1=0\,,
%\label{eq014}\\
%&\textstyle
%W_1=F W_0-\partial_\varphi W_2\,,
%\label{eq015}\\
%&\textstyle
%W_{2m}=(2m-1)\sigma^2W_{2(m-1)}\,,
%\label{eq016}\\
%&\textstyle
%W_{2m+1}=2m\sigma^2W_{2m-1}+F W_{2m}-\frac{1}{2m+1}\partial_\varphi W_{2(m+1)}\,.
%\label{eq017}
%\end{align}
%Eq.~(\ref{eq016}) yields
%\[
%W_{2m}=(2m-1)!!\,\sigma^{2m}W_0\,;
%\]
%from~(\ref{eq017})
%\[
%W_{2m+1}=2m\sigma^2W_{2m-1}+(2m-1)!!\,\sigma^{2m}(F-\sigma^2\partial_\varphi)W_0\,.
%\]
%With $W_2=\sigma^2W_0$, Eqs.~(\ref{eq014})--(\ref{eq015}) yield
%\[
%W_1=(F-\sigma^2\partial_\varphi)W_0\,,
%\]
%and
\begin{equation}
\partial_t W_0+\partial_\varphi(F W_0)-\sigma^2\partial_\varphi^2W_0=0\,.
\label{eq018}
\end{equation}
Thus, we obtain a conventional Fokker--Planck equation for $W_0$, and all $W_{n\ge1}$ are trivially determined by $W_0$ (see~\cite{Goldobin-Klimenko-2020} for detailed equations).
% For deriving Eq.~(\ref{eq018}), it is sufficient to use Eqs.~(\ref{eq014})--(\ref{eq016}).

%\subsubsection{Corrected Smoluchowski equation ($\mu^1$-correction)}
%Let us derive the $\mu^1$-correction to Eq.~(\ref{eq018}).
Keeping $\mu^1$-corrections for $W_0$, one can find from the infinite equation system~(\ref{eq010})--(\ref{eq013})~\cite{Goldobin-Klimenko-2020};
%\begin{align}
%&\textstyle
%\partial_t W_0+\partial_\varphi W_1=0\,,
%\label{eq019}\\
%&\textstyle
%W_1=F W_0-\partial_\varphi W_2-\mu\partial_t W_1\,,
%\label{eq020}\\
%&\textstyle
%W_2=\sigma^2W_0+\mu\left[-\frac12\partial_t W_2 +F W_1-\frac12\partial_\varphi W_3\right]\,,
%\label{eq021}\\
%&\textstyle
%W_3=2\sigma^2W_1+F W_2-\frac13\partial_\varphi W_4+\mathcal{O}(\mu)\,,
%\label{eq022}\\
%&\textstyle
%W_4=3\sigma^2W_2+\mathcal{O}(\mu)\,.
%\label{eq023}
%\end{align}
%
%Starting with substitution of $W_4$ into the expression for $W_3$, one can find step-by-step:
%\begin{align}
%&\textstyle
%W_3=2\sigma^2W_1+F W_2-\sigma^2\partial_\varphi W_2+\mathcal{O}(\mu)\,,
%\nonumber\\
%&\textstyle
%W_2=\sigma^2W_0+\mu\big[-\frac{\sigma^2}{2}\partial_t W_0
% +F(F W_0-\sigma^2\partial_\varphi W_0)
%\nonumber\\
%&\textstyle\qquad\quad
% -\frac{\sigma^2}{2}\partial_\varphi(F W_0)+\frac{\sigma^4}{2}\partial_\varphi^2W_0
%\nonumber\\
%&\textstyle\qquad\quad
% -\sigma^2\partial_\varphi(F W_0-\sigma^2\partial_\varphi W_0)\big]+\mathcal{O}(\mu^2)\,,
%\nonumber\\
%&W_1=F W_0-\sigma^2\partial_\varphi W_0+\mu\big[-(\partial_tF
% +F\partial_\varphi F)W_0
%\nonumber\\
%&\textstyle\qquad\quad
% +\sigma^2(\partial_\varphi F)\partial_\varphi W_0\big]+\mathcal{O}(\mu^2)\,.
%\nonumber
%\end{align}
%
%Finally, to the $\mu^1$-order,
\begin{align}
\partial_tW_0+\partial_\varphi\big[(F
 -\mu(\partial_tF+F\partial_\varphi F))\,W_0\big]
%\quad
%\nonumber\\
% {}
-\sigma^2\partial_\varphi\big[(1-\mu\partial_\varphi F)\,\partial_\varphi W_0\big]=0\,.
\label{eq024}
\end{align}
This is the corrected Smoluchowski equation~\cite{Gardiner-1983,Wilemski-1976}.
%The effective Langevin equation (Stratonovich form) of Fokker--Planck equation~(\ref{eq024}) is
%\[
%\textstyle
%\dot\varphi=F-\mu\left(\partial_t+F\partial_\varphi +\frac{\sigma^2}{2}\partial_\varphi^2\right)F
% +\sigma\sqrt{1-\mu\partial_\varphi F}\,\xi(t)\,.
%\]

%\subsubsection{High-order of approximations}
The conventional adiabatic elimination of a fast variable requires first three moments $w_0$, $w_1$, $w_2$; the first correction for small $\mu$ requires $w_3$ and $w_4$. Running equation system~(\ref{eq003})--(\ref{eq005}) for $w_0$, $w_1$, ..., $w_{2m+2}$ with formal closure $w_{2m+3}=0$ yields the order of accuracy $\mu^m$.

\subsection{Cumulant formalism}
The equation system for $w_n$,
\begin{align}
\textstyle
nw_n+\mu\partial_t w_n=nF w_{n-1}-\mu\partial_\varphi w_{n+1}
%\qquad
%\nonumber\\
%\textstyle
%{}
+n(n-1)\frac{\sigma^2}{\mu}w_{n-2}\,,
\nonumber
\end{align}
in terms of $f(s,\varphi)=\sum_{n=0}^{+\infty}w_n\frac{s^n}{n!}$ acquires the following form:
\begin{align}
\textstyle
(s\partial_s+\mu\partial_t)f=(sF-\mu\partial_s\partial_\varphi
 +s^2\frac{\sigma^2}{\mu})f\,.
\nonumber
\end{align}
For $\phi=\ln f$, $\partial f=f\partial\phi$,
\begin{align}
\textstyle
(s\partial_s+\mu\partial_t)\phi=sF+s^2\frac{\sigma^2}{\mu}
 -\mu[\partial_s\partial_\varphi\phi+(\partial_s\phi)(\partial_\varphi\phi)]\,.
\nonumber
\end{align}
With $\phi=\sum_{n=0}^{+\infty}K_n\frac{s^n}{n!}$,
\begin{align}
\textstyle
\mu\partial_tK_0&\textstyle=-\mu[\partial_\varphi K_1+K_1\partial_\varphi K_0]\,,
\label{eq025}\\[5pt]
\textstyle
(n+\mu\partial_t)K_n&\textstyle=F\delta_{1n}+\frac{2\sigma^2}{\mu}\delta_{2n}
 -\mu\big[\partial_\varphi K_{n+1}
%\nonumber\\
%&\textstyle\hspace{-1cm}
 +\sum\limits_{j=0}^{n}\frac{n!}{j!(n-j)!}K_{j+1}\partial_\varphi K_{n-j}\big]
\quad\mbox{ for }n\ge1\,.
\label{eq026}
\end{align}
%For the ease of comparison we introduce $\varkappa_n=\frac{K_n}{n!}$ and recast the latter equation system as
%\begin{align}
%\textstyle
%\mu\partial_t\varkappa_0&\textstyle=-\mu[\partial_\varphi\varkappa_1 +\varkappa_1\partial_\varphi\varkappa_0]\,,
%\qquad
%\label{eq027}\\
%\textstyle
%(n+\mu\partial_t)\varkappa_n&\textstyle=F\delta_{1n}+\frac{\sigma^2}{\mu}\delta_{2n}
% -\mu\Big[(n+1)\partial_\varphi\varkappa_{n+1}
% \nonumber\\
%&\textstyle\quad
%+\sum\limits_{j=1}^{n+1}j\varkappa_{j}\partial_\varphi\varkappa_{n+1-j}\Big]
%\;\mbox{ for }n\ge1\,.
%\label{eq028}
%\end{align}

Notice, with equations~(\ref{eq025})--(\ref{eq026}), the conventional elimination of a fast variable requires the first three cumulants (or $w_0$, $w_1$, $w_2$ with equations~(\ref{eq003})--(\ref{eq005})); the first correction for small $\mu$ requires additionally $w_3$ and $w_4$, while, as was shown in~\cite{Goldobin-Klimenko-2020}, within the framework of a cumulant formalism the same first three equations of~(\ref{eq025})--(\ref{eq026}) are sufficient to obtain
\begin{align}
\textstyle
\partial_tK_0&\textstyle=-(\partial_\varphi+K_0^\prime)\big[F
 -\sigma^2K_0^\prime
 +\mu(\partial_tF+F^\prime F
%\nonumber\\
%&\textstyle\qquad\qquad
 +\sigma^2F^\prime K_0^\prime)\big]
 +\mathcal{O}(\mu^2)\,,
\label{eq031}\\
\textstyle
K_1&\textstyle=F-  \sigma^2K_0^\prime
 -\mu\big(\partial_tF+F^\prime F
%\nonumber\\
%&\textstyle\qquad
 +\sigma^2F^\prime K_0^\prime\big)+\mathcal{O}(\mu^2)\,,
\nonumber\\
\textstyle
K_2&\textstyle=\frac{\sigma^2}{\mu}
 -\sigma^2F^\prime+\sigma^4K_0^{\prime\prime}
+\mathcal{O}(\mu)\,.
%% -\frac{\mu\sigma^2}{2}\big[ -3\partial_t\partial_\varphi K_1
%%\nonumber\\
%%&\textstyle
%% +3\sigma^2\partial_\varphi^3K_1
%% -2(\partial_\varphi K_1)^2
%% \nonumber\\
%%&\textstyle
%% -4K_1\partial_\varphi^2K_1+F\partial_\varphi^2K_1\big] +\mathcal{O}(\mu^2)\,,
%\nonumber\\
%\textstyle
%K_n&\textstyle=(-\sigma^2\partial_\varphi)^{n-1}(F
% -\sigma^2K_0^\prime)+\mathcal{O}(\mu)\;\mbox{ for }n\ge3\,.
\nonumber
\end{align}
%One can see that the self-contained equation for the evolution of $K_0$ is more lengthy than Eq.~(\ref{eq14}) for $w_0$.
We can see that equation~(\ref{eq031}) is equivalent to corrected Smoluchowski equation~(\ref{eq024}), if one substitutes
$K_0=\ln W_0$ and notices that $\partial K_0=W_0^{-1}\partial W_0$\,,
$(\partial_\varphi+K_0^\prime)(\cdot)=W_0^{-1}\partial_\varphi(W_0(\cdot))$\,.

To summarize, cumulant equations~(\ref{eq025})--(\ref{eq026}) for finite $\mu$ are significantly more lengthy, than equations for moments $w_n$. However, the convergence properties of $K_n$ for $\mu\to0$ are better, than that of $w_n$. The adiabatic elimination of velocity in terms of $K_n$ and $w_n$ requires the first three elements. However, the $\mu^{1}$-correction to the Smoluchowski equation in terms of $w_n$ requires 5 terms (see~\cite{Wilemski-1976} for the multiple-dimension case), while in terms of cumulants $K_n$ the same first three elements $K_0$, $K_1$, $K_2$ are sufficient. Generally, for the $\mu^m$-correction one needs the leading order accuracy for $K_{m+1}$, i.e., the first $m+2$ cumulants are required. Meanwhile, in terms of $w_n$ (or $W_n$), one needs the first $2m+3$ moments.

\subsection{Basis of Hermite functions}
A conventional way for handling the fast velocity variable in the Fokker--Planck equation is the employment of the basis of Hermite functions for $v$~\cite{Gardiner-1983,Komarov-Gupta-Pikovsky-2014}.
For operator $\hat{L}_1=\partial_u(u+\partial_u)$---which lies in the foundation of the adiabatic elimination of the velocity in FP equation~(\ref{eq002}) for $\rho(v,\varphi)$\,:
\[
\partial_t\rho=-v\partial_\varphi\rho+\partial_v\left[\frac{1}{\mu}\big(v-F(\varphi,t)\big)\rho\right]
+\frac{\sigma^2}{\mu^2}\partial_v^2\rho
\]
---one can see that $\hat{L}_1h_n(u)=-nH_n(u)$,
\[
h_n(u)=H_n(u)\frac{1}{\sqrt{2\pi}}e^{-u^2/2}\,,
\]
$H_n(u)$ is the $n$-th Hermite polynomial of the order $n$, which obeys
\begin{equation}
H_n^{\prime\prime}-uH_n^\prime=-nH_n\,.
\label{eqH01}
\end{equation}

With normalization condition
\[
\int_{-\infty}^{+\infty}\mathrm{d}u\,h_n(u)h_m(u)=\frac{n!\,\delta_{nm}}{\sqrt{2\pi}}
\]
(which provides $\int_{-\infty}^{+\infty}\mathrm{d}u\,h_0(u)=1$),
the following recursive formulae are valid:
%\begin{equation}
%\textstyle
%H_n^\prime=nH_{n-1}\,,
%\label{eqH02}
%\end{equation}
%\begin{equation}
%\textstyle
%uH_n=nH_{n-1}+H_{n+1}\,.
%\label{eqH03}
%\end{equation}
$H_n^\prime=nH_{n-1}$ and $uH_n=nH_{n-1}+H_{n+1}$.
With %Eqs.~(\ref{eqH02}) and (\ref{eqH03}),
these recursive formulae,
the Fokker--Planck equation~(\ref{eq002}) (see also equation~(4) in~\cite{Komarov-Gupta-Pikovsky-2014}) for
\[
\rho=\sum_{n=0}^\infty \frac{\sigma}{\sqrt{\mu}}\,h_n\!\!\left(\frac{\sqrt{\mu}}{\sigma}v\right)W_n(\varphi,t)
\]
yields
%\begin{align}
%&\textstyle
%\sum\limits_nh_n\dot{W}_n(\varphi,t) =\sum\limits_n\Big[-\frac{\sigma}{\sqrt{\mu}}\left(nh_{n-1}\right.
%\nonumber\\
%&\textstyle\quad
%\left.{}+h_{n+1}\right)\partial_\varphi W_n(\varphi,t)
%  -\frac{n}{\mu}h_nW_n(\varphi,t)
%\nonumber\\
%&\textstyle\qquad
% {}+\frac{F}{\sigma\sqrt{\mu}}h_{n+1}
%W_n(\varphi,t)\Big]\,.
%\nonumber
%\end{align}
%After projecting:
\begin{align}
\dot{W}_0&=-\frac{\sigma}{\sqrt{\mu}}\partial_\varphi W_1\,,
\label{eqH04}\\
\textstyle
\dot{W}_n&=\frac{\sigma}{\sqrt{\mu}}\big((\sigma^{-2}F-\partial_\varphi)W_{n-1}
%\nonumber\\
%&\quad
% {}
 -(n+1)\partial_\varphi W_{n+1}\big)
  -\frac{n}{\mu}W_n\,
\quad
\;\mbox{ for } n\ge1\,.
\label{eqH05}
\end{align}

\subsubsection{Elimination of a fast variable}
For $\mu\ll1$, it is more convenient to rewrite equations~(\ref{eqH04})--(\ref{eqH05}) as
\begin{align}
\dot{W}_0&=-\frac{\sigma}{\sqrt{\mu}}\partial_\varphi W_1\,,
\label{eqH04-mu}\\
\textstyle
W_n&=\frac{\sqrt{\mu}\,\sigma}{n}\big((\sigma^{-2}F-\partial_\varphi)W_{n-1}
%\nonumber\\
%&\quad
% {}
 -(n+1)\partial_\varphi W_{n+1}\big)
  -\frac{\mu}{n}\partial_tW_n
\quad
\;\mbox{ for } n\ge1\,.
\label{eqH05-mu}
\end{align}

With equations~(\ref{eqH04-mu})--(\ref{eqH05-mu}), one finds $W_n\sim\mu^{n/2}$.

Taking the leading order for $W_N$, one has $\mathrm{error}(W_N)\sim\mu^{N/2+1}$, $\mathrm{error}(W_{N-1})\sim\mu^{N/2+1+1/2}$, \dots, $\mathrm{error}(W_1)\sim\mu^{N/2+1+(N-1)/2}$, and $\mathrm{error}(\partial_tW_0)\sim\mu^{N}$. Thus, the truncation after $W_N$ leads to inaccuracy $\sim\mu^{N}$.

\subsection{``Cumulant'' formalism for the Hermite function basis}
Let us construct an analogue of cumulant representation for $v$. For generating function $f(s,\varphi)=\sum_{n=0}^\infty W_n(\varphi)s^n$ (it will be essential below to use $s^n$, but not $s^n/n!$), one finds
\[
%\textstyle
\dot{f}=\frac{\sigma}{\sqrt{\mu}}\big(s(\sigma^{-2}F-\partial_\varphi)f
 -\partial_s\partial_\varphi f\big)-\frac{1}{\mu}s\partial_sf\,.
\]

For $\Phi=\ln f$, $\partial\Phi=\partial f/f$,
\begin{align}
&%\textstyle
\dot{\Phi}=\frac{\sigma}{\sqrt{\mu}}\big(s(\sigma^{-2}F-\partial_\varphi\Phi)
%\nonumber\\
%&\textstyle\qquad\qquad
% {}
 -\partial_s\partial_\varphi\Phi-(\partial_s\Phi)(\partial_\varphi\Phi)\big)
 -\frac{1}{\mu}s\partial_s\Phi\,.
\nonumber
\end{align}
With $\Phi(s,\varphi)=\sum_{n=0}^\infty\varkappa_n(\varphi)s^n$, the latter equation yields
\begin{align}
\textstyle
\dot\varkappa_0
&\textstyle
=-\frac{\sigma}{\sqrt{\mu}}(\partial_\varphi\varkappa_1 +\varkappa_1\partial_\varphi\varkappa_0)\,,
\label{eqH06}
\\
\textstyle
\dot\varkappa_n
&\textstyle
=\frac{\sigma}{\sqrt{\mu}}\big(\sigma^{-2}F\delta_{1n}-\partial_\varphi\varkappa_{n-1}
 -(n+1)\partial_\varphi\varkappa_{n+1}
%\nonumber\\
%&\textstyle\quad
% {}
 -\sum\limits_{n_1+n_2 \atop =n+1}n_1\varkappa_{n_1}\partial_\varphi\varkappa_{n_2}\big)
 -\frac{n}{\mu}\varkappa_n
 \quad
 \mbox{ for }n\ge1\,.
\label{eqH07}
\end{align}
For $\mu\ll1$, it is convenient to recast the latter system as
\begin{align}
\textstyle
\dot\varkappa_0
&\textstyle
=-\frac{\sigma}{\sqrt{\mu}}(\partial_\varphi\varkappa_1 +\varkappa_1\partial_\varphi\varkappa_0)\,,
\label{eqH08}
\\
\textstyle
\varkappa_n
&\textstyle
=\frac{\sqrt{\mu}\,\sigma}{n}\big(\sigma^{-2}F\delta_{1n}-\partial_\varphi\varkappa_{n-1}
 -(n+1)\partial_\varphi\varkappa_{n+1}
%\nonumber\\
%&\textstyle\quad
% {}
 -\sum\limits_{n_1+n_2 \atop =n+1}n_1\varkappa_{n_1}\partial_\varphi\varkappa_{n_2}\big)
 -\frac{\mu}{n}\partial_t\varkappa_n
 \quad
 \mbox{ for }n\ge1\,.
\label{eqH09}
\end{align}
For the $\mu^1$-approximation,
\begin{align}
\textstyle
\dot\varkappa_0
&\textstyle
=-(\varkappa_0^\prime+\partial_\varphi)\big[F-\mu(\partial_t+F^\prime)F
%\nonumber\\
%&\textstyle\qquad\qquad
-\sigma^2(1-\mu F^\prime)\varkappa_0^\prime\big]+\mathcal{O}(\mu^2)\,,
\label{eqH10}
\\
\textstyle
\varkappa_1
&\textstyle
=\sqrt{\mu}\,\sigma\big(\sigma^{-2}F-\partial_\varphi\varkappa_0
 -\mu\big[\sigma^{-2}(\partial_t+F^\prime)F
%\nonumber\\
%&\textstyle\qquad\qquad
% {}
 -F^\prime\varkappa_0^\prime\big]\big)+\mathcal{O}(\mu^{5/2})\,,
\label{eqH11}
\\
\textstyle
\varkappa_2
&\textstyle
=-\frac{\sqrt{\mu}\,\sigma}{2}\partial_\varphi\varkappa_1+\mathcal{O}(\mu^2)\,.
\label{eqH12}
\end{align}
Equation~(\ref{eqH10}) is equivalent to~(\ref{eq024}) (see explanations after equation~(\ref{eq031})).

For system~(\ref{eqH08})--(\ref{eqH09}), $\varkappa_n\sim\mu^{n/2}$; the $\mu^N$-approxi\-ma\-tion requires truncation atfer $\varkappa_{N+1}$.

Here, there seems to be no preference between the $W_n$- and $\varkappa_n$-representations, except the equations in terms of $\varkappa_n$ are more lengthy.

In this subsection, it is essential that the definition of the generating function $f(s,\varphi)$ with $W_ns^n/n!$ is inappropriate, since such a definition leads to the term $\partial_s^{-1}f$ in the governing equation for $f$; this term cannot be represented with simple regular sums in terms of $\varkappa_n$.

%%%%%%%%%%%%%%%%%%%%%%%%%%%%%%%%%%%%%%%%%%%%%%%%%%%%%%%%%%%%%
%%%%%%%%%%%%%%%%%%%%%%%%%%%%%%%%%%%%%%%%%%%%%%%%%%%%%%%%%%%%%
%\begin{figure}[!thb]
%\center{
%\includegraphics[width=0.495\columnwidth]%
% {ABP_-1_+1.eps }
%}
%\caption{Hierarchy of smallness of higher-order elements for active Brownian particles with $\alpha=-1$, $\beta=1$; a series of 50 terms is used, here $||g(\varphi)||\equiv\int_0^{2\pi}|g(\varphi)|\mathrm{d}\varphi$.
%}
%  \label{fig3}
%\end{figure}
%%%%%%%%%%%%%%%%%%%%%%%%%%%%%%%%%%%%%%%%%%%%%%%%%%%%%%%%%%%%%
%%%%%%%%%%%%%%%%%%%%%%%%%%%%%%%%%%%%%%%%%%%%%%%%%%%%%%%%%%%%%

\subsection{Moment and cumulant formalisms for active Brownian particles}
Let us consider the following Langevin equation
\begin{equation}
\mu\ddot\varphi+\alpha\dot\varphi +\beta\dot\varphi^3=F(\varphi,t)+\sigma\xi(t)\,,\quad\mu\ll1\,,
\label{eq-abp-01}
\end{equation}
where $\beta>0$.
This example can be useful only as an illustration since the fluctuation term and the leading dissipation term here are not in concordance with the Fluctuation--Dissipation Theorem.

With the Fokker--Planck equation
\begin{equation}
\partial_t\rho=-v\partial_\varphi\rho +\partial_v\left[\frac{\alpha v+\beta v^3 -F(\varphi,t)}{\mu}\rho\right]
+\frac{\sigma^2}{\mu^2}\partial_v^2\rho\,,
\label{eq-abp-02}
\end{equation}
the moment equation system acquires the following form:
\begin{align}
\textstyle
\alpha nw_n+\beta nw_{n+2}+\mu\partial_t w_n=nF w_{n-1}
%\qquad
%\nonumber\\
%\textstyle
%{}
-\mu\partial_\varphi w_{n+1} +n(n-1)\frac{\sigma^2}{\mu}w_{n-2}\,,
\label{eq-abp-03}
\end{align}
which yields in terms of $f(s,\varphi)=\sum_{n=0}^{+\infty}w_n\frac{s^n}{n!}$:
\begin{align}
\textstyle
(\alpha s\partial_s+\beta s\partial_s^3 +\mu\partial_t)f=(sF-\mu\partial_s\partial_\varphi
 +s^2\frac{\sigma^2}{\mu})f\,.
\nonumber
\end{align}
For $\phi=\ln f$, $\partial f=f\partial\phi$,
\begin{align}
\textstyle
(\alpha s\partial_s +\mu\partial_t)\phi
 +\beta s[\partial_s^3\phi+3\partial_s\phi\partial_s^2\phi +(\partial_s\phi)^3]
%\quad
%\nonumber\\
%\textstyle
 =sF+s^2\frac{\sigma^2}{\mu}
 -\mu[\partial_s\partial_\varphi\phi+(\partial_s\phi)(\partial_\varphi\phi)]\,.
\nonumber
\end{align}
With $\phi=\sum_{n=0}^{+\infty}K_n\frac{s^n}{n!}$,
\begin{align}
\textstyle
&\textstyle
\mu\partial_tK_0=-\mu[\partial_\varphi K_1+K_1\partial_\varphi K_0]\,,
\label{eq-abp-04}\\[5pt]
&\textstyle
(\alpha n+\mu\partial_t)K_n+\beta n\Big[K_{n+2} +3\sum\limits_{j=1}^{n}\frac{(n-1)!}{(j-1)!(n-j)!}K_{j}K_{n+2-j}
%\nonumber\\
%&\textstyle\quad
+\!\!\!\sum\limits_{j_1+j_2+j_3 \atop =n+2}\!\!\frac{(n-1)!}{(j_1-1)!(j_2-1)!(j_3-1)!}K_{j_1}K_{j_2}K_{j_3}\Big]
\nonumber\\
&\qquad
\textstyle=F\delta_{1n}+\frac{2\sigma^2}{\mu}\delta_{2n}
 -\mu\big[\partial_\varphi K_{n+1}
%\nonumber\\
%&\textstyle\quad
 +\sum\limits_{j=0}^{n}\frac{n!}{j!(n-j)!}K_{j+1}\partial_\varphi K_{n-j}\big]
\quad\mbox{ for }n\ge1\,.
\label{eq-abp-05}
\end{align}
%In terms of $\varkappa_n=\frac{K_n}{n!}$,
%\begin{align}
%\textstyle
%\mu\partial_t\varkappa_0&\textstyle=-\mu[\partial_\varphi\varkappa_1 +\varkappa_1\partial_\varphi\varkappa_0]\,,
%\qquad
%\label{eq027}\\
%\textstyle
%(n+\mu\partial_t)\varkappa_n&\textstyle=F\delta_{1n}+\frac{\sigma^2}{\mu}\delta_{2n}
% -\mu\Big[(n+1)\partial_\varphi\varkappa_{n+1}
% \nonumber\\
%&\textstyle\quad
%+\sum\limits_{j=1}^{n+1}j\varkappa_{j}\partial_\varphi\varkappa_{n+1-j}\Big]
%\;\mbox{ for }n\ge1\,.
%\label{eq028}
%\end{align}

For the first five equations of (\ref{eq-abp-04})--(\ref{eq-abp-05}),
\begin{align}
&\textstyle
\partial_tK_0=-\partial_\varphi K_1-K_1\partial_\varphi K_0\,,
\qquad
\nonumber\\
&\textstyle
(\alpha+\mu\partial_t)K_1
+\beta[K_3+3K_1K_2+K_1^3]
% \nonumber\\
%&\textstyle\qquad
 =F -\mu\big[\partial_\varphi K_2
 +K_1\partial_\varphi K_1+K_2\partial_\varphi K_0\big]\,,
\nonumber\\
&\textstyle
(2\alpha+\mu\partial_t)K_2 +2\beta\big[K_4+3(K_1^2+K_1K_3)+3K_1^2K_2\big]
 \nonumber\\
&\textstyle\qquad\quad
 =\frac{2\sigma^2}{\mu} -\mu\big[\partial_\varphi K_3
 +K_1\partial_\varphi K_2+2K_2\partial_\varphi K_1+K_3\partial_\varphi K_0\big]\,,
\nonumber\\
&\textstyle
(3\alpha+\mu\partial_t)K_3
 +3\beta\big[K_5+3(3K_3K_2+K_1K_4)
 +3K_1^2K_3+6K_2^2K_1\big]
\nonumber\\
&\textstyle\qquad\quad
 =-\mu\big[\partial_\varphi K_4 +K_1\partial_\varphi K_3
% \nonumber\\
%&\textstyle\qquad\qquad
+3K_2\partial_\varphi K_2+3K_3\partial_\varphi K_1+K_4\partial_\varphi K_0\big]\,,
\nonumber\\
&\textstyle
(4\alpha+\mu\partial_t)K_4+4\beta\big[K_6+3(4K_4K_2+3K_3^2+K_1K_5)
%\nonumber\\
%&\textstyle\qquad\qquad
 +6K_2^3+18K_1K_2K_3+3K_1^2K_4\big]
 \nonumber\\
&\textstyle\qquad\quad
 =-\mu\big[\partial_\varphi K_5 +K_1\partial_\varphi K_4 +4K_2\partial_\varphi K_3
% \nonumber\\
%&\textstyle\qquad\qquad
+6K_3\partial_\varphi K_2+4K_4\partial_\varphi K_1+K_5\partial_\varphi K_0\big]\,,
\nonumber\\
&\qquad\qquad\dots\;.
\label{eq-abp-06}
\end{align}

The inspection of equation system~(\ref{eq-abp-06}) reveals the following scaling properties of $K_n$ with respect to $\mu$:
\[
K_n\sim\left\{
\begin{array}{cl}
\displaystyle
  \mu^{-\frac{n}{4}} & \mbox{  for even }$n$\,, \\
\displaystyle
  \mu^{\frac{3}{4}-\frac{n}{4}} & \mbox{  for odd }$n$\,.
\end{array}\right.
\]
With such scaling properties the $\beta$- and $\sigma^2$-terms for even $n$ in equation system~(\ref{eq-abp-06}) dominate and one cannot truncate the equation chain without affecting the leading order with respect to $\mu$. Moreover, one faces similar issue with even $n$; which is coupled with the $F$-term in the leading order. Thus, the calculations in the leading order require $\beta$-, $F$- and $\sigma^2$-terms and these calculations in terms of $K_n$ (or $w_n$) are extremely challenging. This problem can be more efficiently solved with the Fokker--Planck equation~(\ref{eq-abp-02}) where all terms without $\beta$, $F$ or $\sigma^2$ are dropped. One finds
\[
\rho=C(\varphi)e^{\frac{\mu}{\sigma^2}(-\frac{\beta v^4}{4}+F v)}+\dots\,,
\]
where $\dots$ stand for higher-order corrections. After laborious but straightforward calculations one can obtain:
\begin{align}
  w_{2n+1}&\approx\frac{4F}{\beta}\frac{\Gamma(\frac{n}{2}+\frac34)}{\Gamma(\frac14)} \left(\frac{2\sigma}{\sqrt{\beta\mu}}\right)^{n-1}w_0\,,
\label{eq-abp-07}\\
  w_{2n}&\approx\frac{\Gamma(\frac{n}{2}+\frac14)}{\Gamma(\frac14)} \left(\frac{2\sigma}{\sqrt{\beta\mu}}\right)^{n}w_0\,.
\label{eq-abp-08}
\end{align}
Corresponding cumulants:
\begin{align}
&\textstyle
K_0=\ln w_0\,,\qquad
K_2\approx\frac{\Gamma(\frac34)}{\Gamma(\frac14)}
  \frac{2\sigma}{\sqrt{\beta\mu}}\,,
%\nonumber\\
%&
\qquad
K_4\approx-\left(3\left[\frac{\Gamma(\frac34)}{\Gamma(\frac14)}\right]^2-\frac14\right)
  \frac{4\sigma^2}{\beta\mu}\,,\quad
\nonumber\\
&\textstyle
K_6\approx3\frac{\Gamma(\frac34)}{\Gamma(\frac14)} \left(10\left[\frac{\Gamma(\frac34)}{\Gamma(\frac14)}\right]^2-1\right)
  \left(\frac{2\sigma}{\sqrt{\beta\mu}}\right)^3\,,\,\dots\,,
\nonumber\\
&\textstyle
K_1\approx\frac{4F}{\beta}\frac{\Gamma(\frac34)}{\Gamma(\frac14)} \frac{\sqrt{\beta\mu}}{2\sigma}\,,
%\nonumber\\
%&
\qquad
K_3\approx-\frac{4F}{\beta}\left(3\left[\frac{\Gamma(\frac34)}{\Gamma(\frac14)}\right]^2 -\frac14\right)\,,
\nonumber\\
&\textstyle
K_5\approx3\frac{4F}{\beta}\frac{\Gamma(\frac34)}{\Gamma(\frac14)} \left(10\left[\frac{\Gamma(\frac34)}{\Gamma(\frac14)}\right]^2 -1\right)\frac{2\sigma}{\sqrt{\beta\mu}}\,,\,\dots\,.
\nonumber
\end{align}

Obviously, the cumulant representation can be beneficiary mainly for the systems where the distribution of the fast variable is close to the Gaussian distribution. An example of the latter is the case of passive Brownian particles, where the Fluctuation--Dissipation theorem requires the Gaussian distribution for the unperturbed state, and can be often relevant for active Brownian particles, where the leading dissipation term is in concordance with the fluctuations-term.

\section{Conclusion}
The four analyzed formalisms for elimination of a fast variable (velocity) yield a comparable accuracy for the same strong order of accuracy with respect to the inertia parameter $\mu$ (mass). However, for the moment formalism employing $w_n(\varphi)=\int v^n\rho(v,\varphi)\mathrm{d}v$, the strong order $\mu^m$ requires $2m+3$ equations (from the order 0 to the order $2m+2$); for corresponding cumulants $K_n$, only $m+2$ equations are required (from 0th to $(m+1)$th orders); for the Hermite function basis and its formal `cumulant' version, the same $m+2$ equations are required.

For the case of active Brownian particles one cannot employ the Hermite function basis, while one can still use the moment or cumulant formalisms. Practical implementation of these formalisms for numerical simulation can be efficiently performed with employment of a modification of the exponential time differencing method~\cite{Permyakova-Goldobin-2020}.

\ack{The work of EVP and LSK on the case of active Brownian particles was supported by the Russian Science Foundation (grant no.\ 19-42-04120).
The work of IVT and DSG on the case of passive Brownian particles was supported by the Russian Science Foundation (grant no.\ 19-12-00367).}

\end{document}